\documentclass[twoside,11pt]{article}
\usepackage{graphicx}
\usepackage{bm}
\usepackage{booktabs}
\usepackage[round]{natbib}
\usepackage[margin=1in]{geometry}
\usepackage[usenames]{color}
\usepackage{float}
\usepackage{amsfonts, amssymb, amsmath, amsthm, multicol}
\usepackage[colorlinks=true, pdfstartview=FitV, linkcolor=blue,
citecolor=blue, urlcolor=blue]{hyperref}
\usepackage[usenames]{color}
\usepackage[T1]{fontenc}
\usepackage[utf8]{inputenc}
\usepackage{lmodern}
\definecolor{Red}{rgb}{1,0.05,0}
\definecolor{Grn}{rgb}{0.1,0.7,0.1}
\definecolor{Blu}{rgb}{0.1,0.1,0.6}
\definecolor{Org}{rgb}{1,0.45,0}
\definecolor{Vio}{rgb}{0.6578,0,0.9478}
\definecolor{Mag}{rgb}{1,0.2,0.3}
\definecolor{Cya}{rgb}{0,0.7,0.8}        
\definecolor{Tea}{rgb}{0,0.5,0.5}        
\definecolor{Nav}{rgb}{0,0,0.5}          
\definecolor{Sky}{rgb}{0.4,0.7,1}        
\definecolor{Pur}{rgb}{0.5,0,0.5}        
\definecolor{Pin}{rgb}{1,0.4,0.7}        
\definecolor{Brn}{rgb}{0.6,0.3,0.1}      
\usepackage{authblk}
 \usepackage{booktabs}

\title{Hydrodynamically engineered Indigenous arrows skip on water for waterfowl hunting}

\author[1]{ {{\color{Grn}Junrong Zhang}}}
\author[2]{{\color{Org} {Farrukh Kamoliddinov}}}
\author[1]{{\color{Vio} Thomas Yang}}
\author[1]{{\color{Blu} Jeff Tang}}
\author[1]{{\color{Cya} Tino Liang}}
\author[1]{{\color{Brn} Ethan Tam}}
\author[1]{{\color{Tea} Edam Jin}}
\author[2]{{\color{Mag} {Tadd Truscott}}}
\author[1]{{\color{Red} Zhao Pan}\thanks{To whom correspondence may be addressed: tadd.truscott@kaust.edu.sa; zhao.pan@uwaterloo.ca}}
\affil[1]{University of Waterloo, Department of Mechanical and Mechatronics Engineering, Waterloo, Ontario, N2L 3G1, Canada}
\affil[2]{Mechanical Engineering, Physical Sciences and Engineering Division, King Abdullah University of Science and Technology (KAUST), 
Thuwal, 23955-6900, Saudi Arabia. }

\date{\today}

\begin{document}

\maketitle \vspace{ 0 cm}

\begin{abstract}  
Across the Northern Hemisphere, Indigenous hunters developed arrows capable of skipping across the water surface to strike waterfowl. Archaeological and ethnographic records reveal remarkably similar projectile designs spanning millennia and geographically distant cultures, suggesting a convergent technological solution. Despite extensive study of water-entry dynamics, the physical principles underlying this behaviour remain poorly understood. 
Here we show that successful water-skipping arises from a small set of coupled geometric and dynamical parameters that define a bounded operational regime separating rebound, plunging, and overshoot. Using a combination of controlled experiments, hydrodynamic modeling, and historical reconstruction, we demonstrate that reconstructed arrow designs from independent cultures consistently fall within this predicted regime.
These results demonstrate that Indigenous technologies were effectively tuned to satisfy the hydrodynamic constraints governing controlled skipping, providing evidence of convergent optimization in human-engineered systems. More broadly, our results suggest that material culture encodes physical knowledge that formal science is only beginning to articulate, and that the archaeological record and Indigenous culture may be an underexplored archive of empirical discovery.
\end{abstract}

Across much of the Northern Hemisphere, hunters have long used arrows and spears designed to skip across the surface of water to strike waterfowl~\citep{vilkuna1950obugrischen, nelson1900eskimo}. 
Ethnographic accounts describe projectiles that, when launched at shallow angles, ricochet repeatedly along the water surface before striking birds resting on the water (Fig.~\ref{fig: introduction}(a,b)). 
By travelling close to the surface and across a broad trajectory, these projectiles allow hunters to target flocks along a line rather than a single point, greatly increasing the likelihood of a successful strike \citep{nelson1900eskimo,vilkuna1950obugrischen}.

Archaeological and historical records \textcolor{black}{in multiple languages} reveal that remarkably similar projectile designs appear across widely separated cultures and time periods. 
Water-skipping arrows and spears are documented among Inuit hunters in Arctic North America, in northern Eurasian hunting traditions, and in imperial Chinese archery practices. 
Archaeological finds indicate that related designs may date back to the Early Mesolithic \citep{vilkuna1950obugrischen,nelson1900eskimo,murdoch1892,zhilin2015early,gross2019working,huangchao_lqts,pilivciauskas2020fishing,joona2011rautoisia,selby2008rediscovering,egede1818greenland,pfeifer2021bows,kylstra1977continuity}, suggesting that this design principle has persisted for several millennia (Fig.~\ref{fig: introduction}(c,d) and Supplementary Information for a comprehensive summary of historical accounts). 
\textcolor{black}{Notably, knowledge of these techniques appears to have been transmitted largely through oral and folkloric traditions rather than formal record, and in many communities this knowledge is now at risk of being lost.}
Despite differences in materials and cultural context, these projectiles share a common geometric feature: an enlarged blunt nose positioned near the arrowhead, often carved from wood, bone, or ivory (Fig.~\ref{fig: introduction}(e--i)). 
The consistency of this design across cultures suggests a technological solution that was \textcolor{black}{independently and repeatedly reinvented or spread through cultural diffusion}, and refined over millennia.

Yet, it remains poorly understood whether these projectiles function as intended, and what physical mechanisms govern their ability to skip across water.
While the hydrodynamics of skipping stones and other ricocheting bodies have been studied extensively \citep{belden2016elastic,rosellini2005skipping,clanet2004secrets} the design rules governing water-skipping hunting projectiles have not been investigated. In particular, it is unclear how projectile geometry and launch conditions interact to produce controlled ricochets that remain close to the water surface while maintaining sufficient forward momentum to strike a target. 

Here we combine historical synthesis, controlled experiments, and hydrodynamic modeling to identify the mechanisms that enable water-skipping arrows to function. We show that successful skipping is governed by a small set of coupled geometric and operational parameters that determine whether a projectile rebounds, plunges into the water, or overshoots its target. 
Remarkably, reconstructed projectile geometries from archaeological and ethnographic records fall within the predicted operational envelope, suggesting that Indigenous hunting technologies converged on hydrodynamically robust designs. The results reveal how a widespread hunting technology exploits fundamental principles of water-entry hydrodynamics and illustrate how empirical engineering traditions can encode sophisticated physical design rules.

\subsection*{Global record of water-skipping arrows}

\begin{figure}
\centering
\includegraphics[width=.94\linewidth]{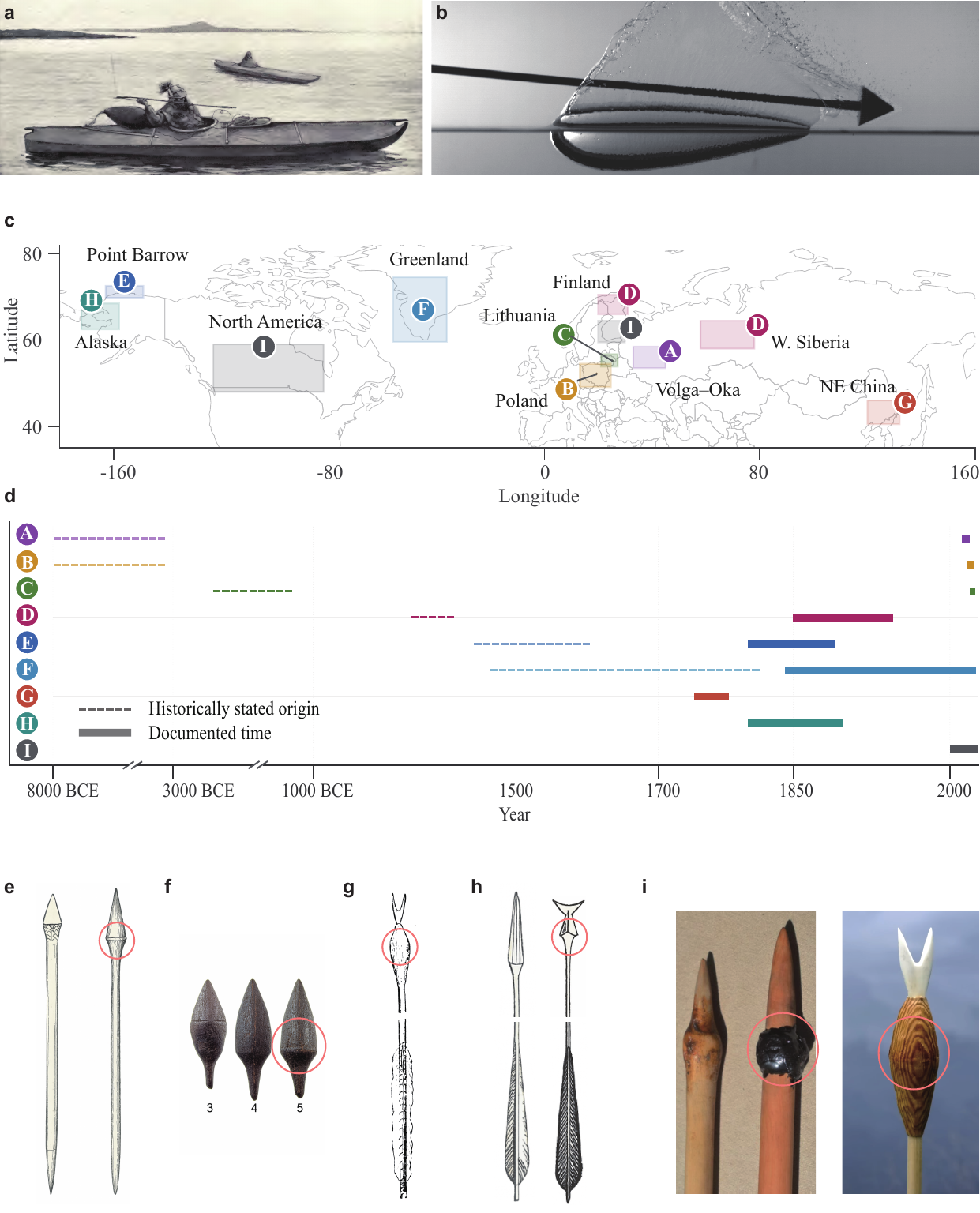}
\caption{\textbf{Global distribution, historical use and convergent design of water-skipping hunting projectiles.}}
    \label{fig: introduction}
\end{figure}

Ethnographic and archaeological records document a distinctive class of hunting projectiles designed to skip across the surface of water before striking waterfowl. Similar projectile forms appear across widely separated cultures as summarized in Fig.~\ref{fig: introduction} including: Inuit bird spears and arrows (Fig.~\ref{fig: introduction}a) \citep{nelson1900eskimo,murdoch1892}, forked or thickened arrows described in northern Eurasian hunting traditions (Fig.~\ref{fig: introduction}(e--g) \citep{zhilin2015early,gross2019working,vilkuna1950obugrischen}), and specialized water arrows recorded in Qing dynasty archery manuals (Fig.~\ref{fig: introduction}h) \cite{huangchao_lqts}. 
Modern reconstructions and experimental replicas further illustrate this behaviour in practice (Fig.~\ref{fig: introduction}i)\citep{woods2015water_skipping_arrow,sami_bouncing_arrows}.

Despite differences in materials, specific design, and cultural context, these projectiles share a consistent geometric feature. 
Many examples incorporate an enlarged blunt structure positioned near the arrowhead (Fig.~\ref{fig: introduction}(e--i)). 
Historical descriptions emphasize the functional importance of this feature: when launched at shallow angles, they ricochet across the water surface and travel close to the surface before striking birds resting on the water, rather than plunging into it \citep{nelson1900eskimo,vilkuna1950obugrischen}. 
The recurrence of this geometry across cultures suggests a design that enhances hydrodynamic rebound upon impact with the water surface is central to the function of these projectiles.

\clearpage
\noindent\textbf{a},~A historical drawing shows the use of a water-skipping spear for waterfowl hunting \citep{nelson1900eskimo}. \textbf{b},~Laboratory observations of water-skipping arrows captured in flight after water rebound. 
\textbf{c},~Geographic distribution of documented water-skipping arrows and spears across the Northern Hemisphere, together with a timeline \textbf{d},~indicating documented records (solid bars), and historically stated earlier origins where available (dashed bars). Dates are shown in calendar years (BCE/CE). These projectiles appear across geographically distant cultures and persist over millennia.
Representative examples from \textbf{e},~Mesolithic biconical arrowheads from the Volga–Oka interfluve \citep{zhilin2015early},  
\textbf{f},~bone and antler arrowheads from Kaltanėnai, Lithuania~\citep{pilivciauskas2020fishing},  
\textbf{g},~Ob-Ugric forked or barbed arrows designed to skim across water~\citep{vilkuna1950obugrischen, joona2011rautoisia},  
\textbf{h}, eighteenth-century Chinese water arrows with pearwood nose~\citep{huangchao_lqts}, and  
\textbf{i},~modern reconstructions and experimental replicas~\citep{woods2015water_skipping_arrow, sami_bouncing_arrows}. Despite their independent origins, these projectiles share a common morphological feature: an enlarged blunt nose that increases wetted area during water impact \textcolor{black}{(circled red)}. 
The recurrence of similar projectile geometries across disparate cultures suggests a convergent technological solution governed by underlying hydrodynamic principles.

\subsection*{Regimes of arrow skipping dynamics}

To characterize the dynamics of water-skipping projectiles, we conducted controlled experiments across a range of launch angles, velocities, and projectile geometries. High-speed imaging reveals that the impact behaviour falls into four distinct regimes: (i) 
surface \emph{skipping} 
, (ii) immediate \emph{plunging} into the water due to improper operation and/or design, and (iii) \emph{overshoot} in which the projectile rebounds over the waterfowl.

\begin{figure*}
\centering
\includegraphics[width=1\linewidth]{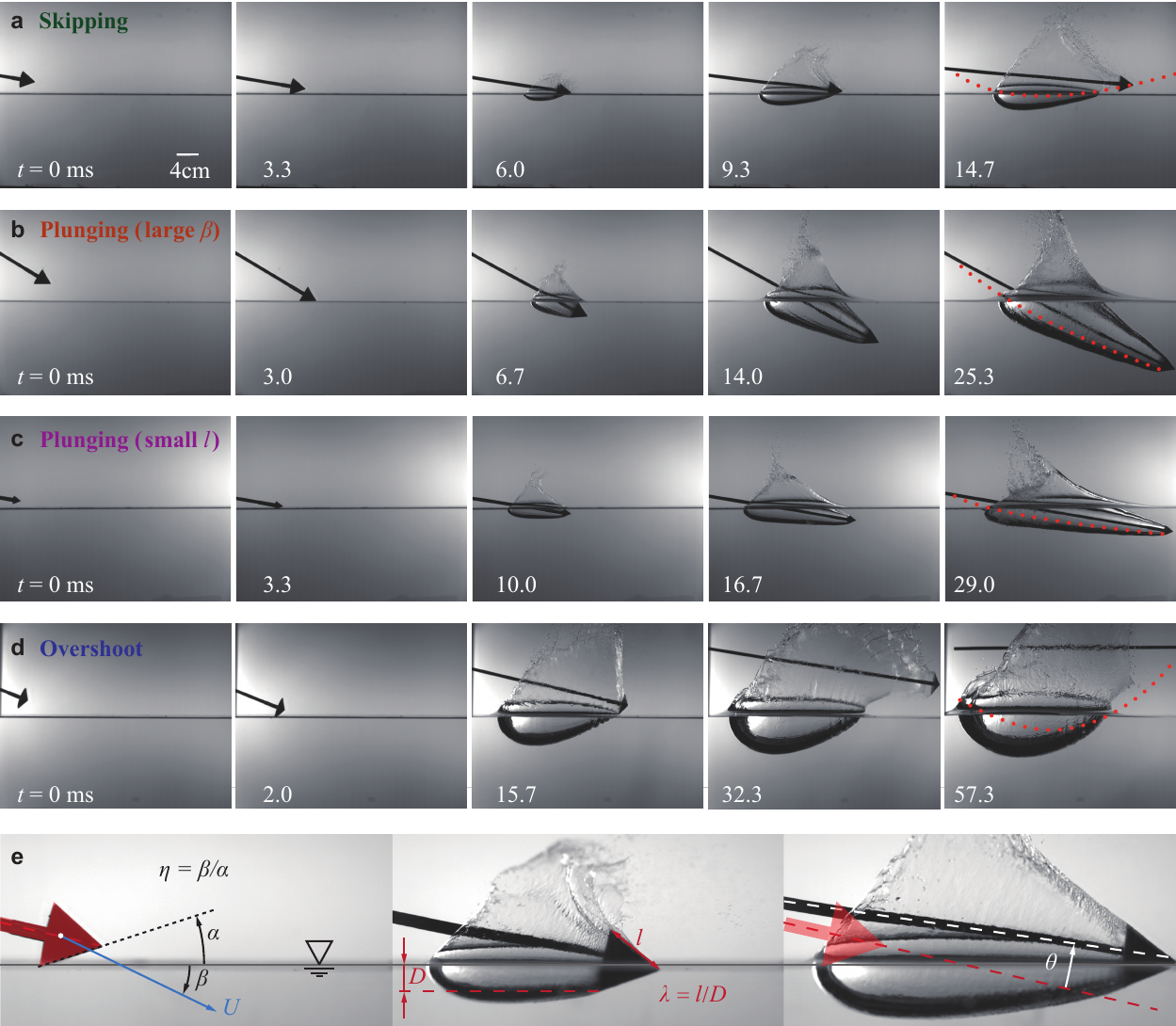}
\caption{\textbf{Water-entry and skipping dynamics.} 
\textbf{a}, Sustained skipping: the projectile impacts the water surface at a shallow angle, generating hydrodynamic lift that redirects its trajectory upward, allowing it to remain near the surface. 
\textbf{b}, Plunging (large $\beta$): the projectile fails to skip due to a steep trajectory.
\textbf{c}, Plunging (small $l$): insufficient lift is generated during impact due to a small nose, and the projectile submerges without rebound.
\textbf{d}, Overshoot: excessive lift produces a steep exit trajectory, causing the projectile to leave the surface and overshoot a near-surface target. 
\textcolor{black}{Trajectories of the arrowhead are marked by the red dotted lines in \textbf{a}--\textbf{d}}. 
\textbf{e}, Schematic illustration of the hydrodynamic mechanism. During shallow-angle impact, pressure develops beneath the nose, generating lift that redirects the trajectory. Variations in impact angle and geometry determine whether lift is insufficient (plunging), sufficient (skipping), or excessive (overshoot).}
    \label{fig: arrow cases}
\end{figure*}

In the skipping regime, the projectile undergoes one or more shallow-angle impacts with the water surface, remaining close to the surface and travelling forward along a near-horizontal trajectory (Fig.~\ref{fig: arrow cases}a). This behaviour occurs when the impact generates sufficient hydrodynamic lift to redirect the projectile upward before it becomes fully submerged (i.e., a shallow trajectory angle ($\beta$) combined with a sufficiently large attack angle ($\alpha$)). 

In contrast, when the trajectory angle is too steep or the impact velocity is insufficient, the projectile plunges into the water without rebounding (Fig.~\ref{fig: arrow cases}b). Projectiles with insufficient nose size ($l$) fail to generate adequate lift even under otherwise favourable conditions and also plunge  (Fig.~\ref{fig: arrow cases}c).

A second failure mode arises when the projectile rebounds too high. For large trajectory angles ($\beta$), the impulsive lift generated during impact produces a steep exit trajectory, causing the projectile to overshoot targets located near the water surface (Fig.~\ref{fig: arrow cases}d). 

These observations indicate that successful water skipping requires a balance between the arrow speed and angle, and projectile geometry. In particular, the ability to generate sufficient lift during impact, while maintaining a shallow exit trajectory, appears to govern whether a projectile can sustain controlled skipping motion for efficient waterfowl hunting.

\subsection*{Hydrodynamic skipping mechanism}

The observed transition between skipping, plunging, and overshoot arises from the balance of hydrodynamic forces acting during water impact. When a projectile strikes the water surface at a shallow angle, pressure builds beneath the nose, generating an upward lift force that can redirect the trajectory before the arrow becomes fully submerged. Whether the projectile rebounds or continues to plunge depends on the interplay between this lift, the projectile inertia, and orientation during impact.

To formalize this behaviour, we model the arrow as a rigid body with a triangular nose interacting with the water surface (Fig.~\ref{fig: arrow cases}e). Following ~\cite{rosellini2005skipping}, aerodynamic and gravitational forces are negligible during the short duration of water contact, and the dynamics are dominated by hydrodynamic loading.
The nondimensional form of the vertical motion of the nose during water contact is governed by
\begin{equation}\label{eq:main_delta}
\ddot{\delta} = -\delta\left(1 - \eta\dot{\delta} + \frac{\theta}{\alpha}\right),
\end{equation}
whereas the airborne phase obeys
\begin{equation}\label{eq:main_air}
\ddot{\delta} = -\varrho .
\end{equation}
Here, $\delta$ denotes the nondimensional draught of the arrowhead, normalized by its maximum draught $D$; $\eta = \beta/\alpha$ is the ratio of trajectory angle to attack angle; $\theta$ represents the rotation of the arrow accumulated during water contact; and $\varrho$ is the nondimensional gravitational acceleration, normalized by the vertical acceleration at the instant of maximum draught.
The derivation of Eqs.~\eqref{eq:main_delta} and \eqref{eq:main_air} can be found in the Supplementary Information. 
\textcolor{black}{These equations show that the skipping dynamics are governed by a small set of coupled parameters describing the impact angle, trajectory, and projectile geometry. In particular, successful skipping for hunting requires satisfying physical constraints that enable rebound while keeping the trajectory near the surface without excessive lift.}

\subsection*{Governing conditions and regime diagram for water-skipping}


The model reveals that successful water skipping requires the simultaneous satisfaction of three physical conditions: (i) sufficient lift to initiate rebound, (ii) adequate nose size to sustain hydrodynamic loading during impact, and (iii) controlled exit trajectories that avoid overshooting the target.

The first condition ensures that the projectile generates sufficient lift at maximum draught to reverse its vertical motion. From Eq.~\eqref{eq:main_delta}, this requires
\begin{equation}
0< \textcolor{black}{\eta \equiv \frac{\beta}{\alpha} }< 1 + \frac{\theta}{\alpha}.
\label{eq: eta small}
\end{equation}
The lower bound $0<\eta$ corresponds to a positive attack angle ($\alpha>0$), while the upper bound defines the maximum $\eta$ for which the lift generated at maximum draught is sufficient to initiate rebound. As $\eta$ increases, the projectile remains submerged longer and rebound is suppressed, consistent with experimental observations (Fig.~\ref{fig: arrow cases}b).

The second condition arises from the finite size of the nose, which limits the maximum hydrodynamic lift. When the nose is too small, the upper surface is easily wetted during full submergence, preventing further lift generation and leading to continued plunging. 
Modelling this effect yields a lower bound on the nondimensional nose size
\begin{equation}
\textcolor{black}{\lambda \equiv \frac{l}{D}} > \frac{1}{\sin(\alpha + \theta)\sqrt{1 + \theta/\alpha}},
\label{eq: lambda criterion}
\end{equation}
\textcolor{black}{where $l$ is the slant height of the nose.}
If this condition is not satisfied, the projectile cannot sustain a complete rebound cycle (Fig.~\ref{fig: arrow cases}c).

The third condition constrains the exit trajectory following the rebound. Excessive lift produces a steep exit angle, causing the projectile to overshoot targets located near the water surface. Requiring that the maximum rebound height remain below a target height $H$ yields a lower bound on $\eta$:
\begin{equation}
\textcolor{black}{\eta \equiv \frac{\beta}{\alpha} } > \hat{\eta}_H =  \frac{3\omega^2}{2}\left(1 - \sqrt{\frac{2gH}{v_0^2}}\right),
\label{eq: eta large}
\end{equation}
where $\hat{\eta}_H$ is a critical $\eta$ for a bird of given height $H$, $\omega^2 = 1 + \theta/\alpha$, and $v_0$ is the vertical speed of the arrow at the moment of water contact, and $g$ is the gravitational acceleration. 
This condition links projectile dynamics directly to the size of the target, indicating that operational parameters must be adjusted according to hunting conditions (Fig.~\ref{fig: arrow cases}d).

\begin{figure*}
\centering
\includegraphics[width=1\linewidth]{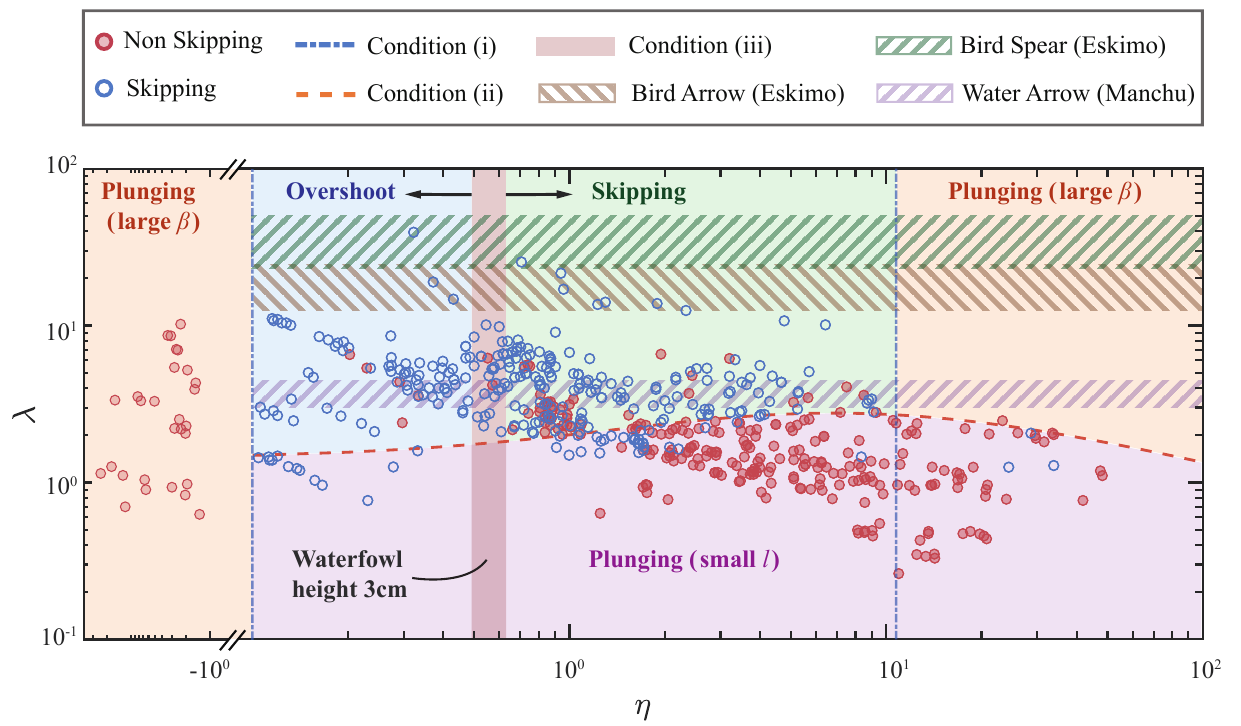}
\caption{\textbf{Regime diagram for water-skipping projectiles and experimental validation in the $\lambda$--$\eta$ plane.}
Vertical chain lines indicate the suitable range of angle ratio $\eta$ for skipping. 
The dashed curve shows the nose-size criterion regarding $\lambda$. 
A vertical semi-transparent red band marks the overshoot constraint for a representative waterfowl height ($H>3$ cm). Effective skips (green-shaded region) occur when the nose is large enough, and the ratio of the trajectory angle to the angle of attack is moderate. 
\textcolor{black}{Experimental data (open blue and filled red circles representing rebound and plunging cases, respectively) reveal that the predicted envelope for skipping is relatively accurate.}
Horizontal hatched bands indicate reconstructed historical arrowhead designs: Manchu water arrows~\citep{huangchao_lqts}, Eskimo bird arrows~\citep{murdoch1892}, and Eskimo bird spears~\citep{nelson1900eskimo}.
These dynamics illustrate how a small set of parameters governs whether the projectile rebounds or submerges during impact. }
\label{fig:4}
\end{figure*}

Together, these conditions define a predictive regime diagram in the $(\lambda,\eta)$ parameter space (Fig.~\ref{fig:4}). The diagram partitions the space into distinct regions corresponding to \textcolor{black}{effective} skipping, plunging, and overshoot.
Experimental measurements collapse onto the theoretically predicted regimes, with successful skipping events falling within the admissible parameter range and failures occurring outside it. 
\textcolor{black}{The three observed failure modes} map directly onto violations of the governing conditions. 
Plunging occurs when insufficient lift is generated to reverse the motion of the projectile, either due to unfavorable impact angles (large $\beta$) or inadequate nose size (small $l$).
Overshoot arises when excessive lift produces a steep exit trajectory that carries the projectile above the target.
\textcolor{black}{Effective skipping for hunting is achieved only when all three conditions are simultaneously satisfied, ensuring both rebound and controlled near-surface motion.} 

Remarkably, reconstructed nose geometries from historical records fall within the predicted feasible region when the trajectory angle and attack angle are properly executed, suggesting that Indigenous projectile designs were tuned to maximize the operational margin for successful skipping, allowing robust performance across a range of launch conditions.

\section*{Acknowledgements}
We thank Sami Maaranen, Peter Dekker, and Prof. Randy Herrmann for valuable discussions. This work was partially supported by the Faculty of Engineering at University of Waterloo and NSERC (CREATE-528202-2019).
\section*{Methods}\label{secA1}

\subsection*{Experimental Setup and Image Processing}
\label{sec: experimental setup}
Experiments were conducted independently at the University of Waterloo and King Abdullah University of Science and Technology (KAUST) using two complementary experimental systems. 
In both cases, projectiles with enlarged conical noses were launched toward a quiescent water surface with controlled impact conditions. 
The impact and rebound dynamics were recorded using \textcolor{black}{calibrated high-speed imaging, from which the impact velocity, trajectory angle, and attack angle were extracted through image processing.}

At University of Waterloo, a laboratory-scale water tank and a custom launcher were used to prescribe the impact angle and velocity of the projectile 
At KAUST, experiments were performed using a recurve bow system capable of achieving higher launch speeds. 
Further details of the experimental setup, projectile geometry, and image-processing procedures are provided in the Supplementary Information.


\subsection*{Numerical Analysis}
We performed numerical analysis of Eqs.~\eqref{eq:main_delta}–\eqref{eq:main_air} to examine how the angle ratio $\eta$ influences the skipping dynamics.
Geometric parameters of the nose and shaft, as well as the arrow mass and initial launch velocity, were fixed to values representative of the physical experiments in our work, while $\eta$ was varied over four representative values ($\eta = 0.1, 1, 10,$ and $100$).
For each simulation solving Eqs.~\eqref{eq:main_delta}–\eqref{eq:main_air}, we applied initial conditions $\delta = 0$ and $\dot\delta = -1$.
The simulation output was represented using the nondimensional arrowhead draught $\delta(t)$ and vertical velocity $\dot{\delta}(t)$ as functions of nondimensional time $\tau$, together with the corresponding portrait on the  $(\dot{\delta},\delta)$ phase plane.
To examine the effect of nose size on rebound dynamics, we analysed numerical solutions of a piecewise dynamical model describing the vertical motion of the nose, which is associated to the development of Eq.~\eqref{eq: lambda criterion}, in the $(\delta,\dot{\delta})$ phase plane. 
Further details and corresponding analysis based on numerical results are provided in the Supplementary Information.

\subsection*{Uncertainty quantification}

Uncertainties in the dimensionless parameters $\lambda$ and $\eta$ were quantified using standard uncertainty propagation
\begin{equation}
\Delta f = \sqrt{
    \sum_{i=1}^{n}
    \left( \frac{\partial f}{\partial x_i} \, \Delta x_i \right)^2 } ,
\end{equation}
where $f$ is a derived quantity and $x_i$ are independent measured variables with uncertainties $\Delta x_i$.

The dimensionless arrowhead size is defined as
\begin{equation}
\lambda = \frac{l}{D}
= l \frac{u}{v_0}
\sqrt{\frac{\rho_w w}{2m}}
\sqrt{\frac{\alpha}{\sin\alpha}},
\end{equation}
and the angle ratio is
\begin{equation}
\eta = \frac{\beta}{\alpha}.
\end{equation}

Geometric dimensions (e.g., arrowhead length $l$ and base diameter $w$) and mass $m$ were measured directly using a calliper (resolution $0.01\,\mathrm{mm}$) and an electronic balance (resolution $0.001\,\mathrm{g}$).
The angular parameters (e.g., $\alpha$, $\beta$, and $\gamma$) were extracted from high-speed videos using PFV4 software.
These measurement uncertainties dominate the propagated uncertainties in $\lambda$ and $\eta$.
\textcolor{black}{Our uncertainty estimation indicates that, for most cases, the relative uncertainty in $\lambda$ is below 20\%, while that in $\eta$ is below 5\%.}
Details of velocity extraction, angular measurement, and full uncertainty propagation are reported in Supplementary Information.

\bibliography{sn-bibliography}
\section*{Data availability}

All data supporting the findings of this study are available within the article and its Supplementary Information.

\end{document}